\documentclass[runningheads]{llncs}
\usepackage{graphicx}
\usepackage{booktabs}
\usepackage[tableposition=above]{caption} 

\usepackage{cite}   
\usepackage[colorinlistoftodos]{todonotes}

\begin{document}

\title{Unsupervised self-organising map of prostate cell Raman spectra shows disease-state subclustering 
}

\titlerunning{Raman Spectra Clustering of Prostate Cells}

\author{Daniel West\inst{1}\orcidID{0000-0002-4522-170X} 
\and
Susan Stepney\inst{1}\orcidID{0000-0003-3146-5401}
\and
Y.~Hancock\inst{2,3}\orcidID{0000-0003-4799-2783}
}

\authorrunning{D. West, S. Stepney, Y. Hancock}

\institute{
Department of Computer Science, University of York, YO10 5GH, UK \and
School of Physics, Engineering and Technology, University of York, YO10 5DD, UK \and
York Biomedical Research Institute, University of York, YO10 5NG}

\maketitle

\begin{abstract}
Prostate cancer is a disease which poses an interesting clinical question: should it be treated? A small subset of prostate cancers are aggressive and require removal and treatment to prevent metastatic spread. However, conventional diagnostics remain challenged to risk-stratify such patients, hence, new methods of approach to biomolecularly subclassify the disease are needed. 
Here we use an unsupervised, self-organising map approach to analyse live-cell Raman spectroscopy data obtained from prostate cell-lines; our aim is to test the feasibility of this method to differentiate, at the single-cell-level, cancer from normal using high-dimensional datasets with minimal preprocessing. 
The results demonstrate not only successful separation of normal prostate and cancer cells, but also a new subclustering of the prostate cancer cell-line into two groups. Initial analysis of the spectra from each of the cancer subclusters demonstrates a differential expression of lipids, which, against the normal control, may be linked to disease-related changes in cellular signalling.  

\end{abstract}

\section{Introduction}
Cancers develop via sequential acquisition of genetic changes within cells, which alter the products the cells produce, their interactions with their environment, and ultimately their behaviour. The earlier these changes are detected, the more chance there is to reverse their effects or to remove the diseased tissue. To assist with this challenge, spectroscopic methods, such as Raman spectroscopy, can be used to provide a non-destructive and label-free means of determining the molecular constitution of cells and environmental impacts on cellular behaviour \cite{Brauchle2013-do}. Raman spectroscopy is a physical technique that can measure down to sub-cellular sampling to produce high-accuracy, molecular-level ``fingerprinting'' of the internal state of cells~\cite{Movasaghi2007-lz}. In its application to the study of cancer, Raman spectroscopy has been shown to specify protein, lipid and DNA/RNA changes that stratify malignant-state characteristics~(e.g. \cite{Potcoava2014-tl}). 

For accurate and statistically robust disease classification, large numbers of high-dimensional Raman spectroscopy data are needed, requiring dimension-reduction methods, such as PCA, to be used. However, PCA, a supervised approach, imposes downstream classification restrictions;  unsupervised methods for dimension reduction and classification may be better suited. Kohonen introduced SOMs in the 1980s as a new form of classification modelling via network topological organisation in which high-dimensional input data are organised by shared statistics and then represented by a two-dimensional array \cite{Kohonen1981-bx,Kohonen1982-mv}. To test their applicability to real-system data, we have performed a feasibility study that uses an unsupervised SOM to classify a prostate cancer cell-line against its normal-state comparison using high-dimensional Raman spectroscopy data. The principles and methods that we demonstrate should also be transferable to other disease-state studies. 

The application of unsupervised machine-learning methods to Raman spectroscopy is still in its infancy, with most research focused on supervised classification methods, predominantly still PCA-LDA \cite{Qi2023-fw}. Biological systems are inherently more complex, and few publications discuss the use of SOMs to analyse samples derived from animal tissue \cite{Brazhe2012-ll, Banbury2019-pa} or human cell lines \cite{Harris2009-ky, Majumdar2020-cl}; all of these methods are either explicitly supervised or do not state whether supervised or unsupervised. Of particular interest in using SOMs is their potential for unsupervised learning and  sub-classification of data. In the field of cancer research, such sub-classification could lead to the discovery of further, stratified disease states, and hence, more targeted and risk-stratified treatment decisions could ensue. The visual SOM output also has benefit for clinicians: it is a readily interpretable and understandable illustration by which patients could then see the position of their own cellular signatures within the complex domain of benign and malignant patterns. 

\section{Background}

\subsection{Domain: Prostate cancer}
This work looks specifically at prostate cancer as a model disease, both because it is common (affecting 1 in 8 men in their lifetime \cite{pcuk}) and because it poses an unusual clinical question amongst cancers: should it be treated? Many forms of prostate cancer are relatively indolent, and only a small subset are highly malignant and aggressive. Decision-making in the management of prostate cancer combines analysis of tissue and blood samples \cite{Epstein2016-yc}, the patient's assessment of symptom severity,
and consideration of what symptoms they could tolerate, whether from the cancer itself or from the side effects of treatment. This process is lengthy and tissue sampling is invasive. Conventional methods of diagnosis, involving imaging and histology, remain challenged in stratifying those individuals who have aggressive forms of the disease: only 15\% of patients require treatment intervention. With the ability of a SOM to sub-stratify complex data, there arises the possibility to improve on these conventional approaches. 

To test the hypothesis that SOMs can successfully cluster high-dimensional medical data, we assess the feasibility of an unsupervised SOM to classify prostate cancer using Raman spectroscopy data collected at the single-cell-level, thereby also testing its ability to sub-stratify the disease-state. Work here uses immortalised cell lines originally derived from human cells extracted and grown \textit{in vitro}. 
These cells were altered to be able to grow independently outside of the host, and are commonly used in research as they form a standard disease model, which allows replication between studies. The two cell lines used here are: PNT2-C2, a cell line derived from healthy prostate epithelium \cite{Berthon1995-ub}; and LNCaP, a cell line derived from a lymph-node deposit of metastatic prostate carcinoma \cite{Horoszewicz1980-lk}. The cells are synchronised to senescence by starvation, such that they are in the same position within the cell cycle when live-cell Raman measurements are taken \cite{Kershaw2017,Cameron2021}.

\subsection{Data: Raman spectroscopy}
Raman spectroscopy involves the inelastic scattering of laser light via photon-molecule interaction resulting in a change in polarisability of the molecule \cite{Ember2017-ue,Ferraro2002-qp}. The measured energy difference between the incident and Raman-scattered photon is typically that which is red-shifted, i.e., Stokes shifted, relative to the Rayleigh elastic-scattering line. Spectral analysis of the Raman-scattered light produces a molecular-scale ``fingerprint'' of the sample. The observed change in photon energy is specific to a particular molecular bond and Raman mode associated with it. As photon energy is directly proportional to frequency, Raman spectra are represented as the integrated Raman-scattering intensity versus wavenumber (i.e., spatial frequency) in units of cm$^{-1}$. 

The Raman data sets used in this investigation correlate to measurements that have $\sim$1 ${\mu}$m, diffraction-limited, spatial resolution, and $\pm$3 cm$^{-1}$ spectral resolution. They comprise high-dimensionality data, which consist of 1056 data points per single-spectrum measurement. The Raman spectra were minimally preprocessed using standard methods of  baseline subtraction, total-area normalisation, and interpolation (see Refs. \cite{Rocha2021-za,Kershaw2017,Cameron2021} and methods details therein). 


\subsection{Analysis: Self-Organising Maps} \label{params}
Self-organising maps (SOMs) are a network of connected threshold-logic units which can assume the topology of an input dataset \cite{Kohonen1982-mv}. Each node is connected to neighbouring nodes to allow competitive feedback when training with a dataset, which enables each node to recognise a unique pattern within the input data \cite{Kohonen1990-sf}. With each round of training, the best matching unit (BMU) is defined as the node which best maps to an input observation pattern, and exerts an effect on its neighbours to bring them closer to itself, thus uncovering clusters of patterns within the dataset.

To allow the SOM the best chance of assuming the topology of the dataset, training data should be normalised. This process improves accuracy of the output map as the normalised input vectors have the same dynamic range \cite{Kohonen2001-al}. For high-dimensionality data such as that of Raman spectroscopy, normalisation of the variance of each dimension across the dataset and a Euclidean measure of distance between vectors is appropriate for most studies \cite{Kohonen2001-al,Kohonen2013-jv}.
The SOM method used here normalises the input spectral data in the intensity direction, and does not alter the data in the wavenumber direction.

To build a SOM, several key parameters of the algorithm must be set to optimise results: map network topology, configuration, and dimensions; neighbourhood function; learning rate; decay function; and maximum iteration number.

\subsubsection{Map Network.}
The nodal array can be arranged in a one-dimensional line or a higher dimensional lattice, with equal or unequal spacing between adjacent nodes. For a two-dimensional sheet of nodes, connections are commonly formed in a rectangular or hexagonal configuration, where each node is connected to four or six neighbouring nodes, respectively. A hexagonal configuration is often preferred for analysis as each node exerts influence over more neighbouring nodes than with a rectangular array \cite{Kohonen2001-al}; a rectangular array output may be easier for non-experts to interpret.

The network must have appropriate size to display data, while not being so large that the learning takes excessive time to run. For a rectangular lattice, taking the $x$ and $y$ dimensions of the map to be the ratio of the two highest eigenvalues of the input data autocorrelation matrix facilitates faster convergence of the model \cite{Kohonen2013-jv}. 
The number of nodes used is a modelling choice, and there is no way to know the ``correct'' number prior to model training. Ultimately, a finer resolution with a larger overall grid size will allow finer resolution of the underlying data, at the cost of computational time. Vesanto \cite{Vesanto2000-rc} suggests the figure $5\sqrt{n}$ nodes, where $n$ is the number of observations in the dataset.  This has become widely used as a starting SOM array size.

\subsubsection{Neighbourhood Function.}
The neighbourhood function, $\sigma(t)$, defines a symmetrical region around the BMU over which the BMU exerts an effect to update the weight of any included nodes and draw them closer to the BMU. The neighbourhood function radius decreases monotonically with each iteration step, $t$.

Using this method, early iteration steps conduct coarse organisation of the input data with a wide neighbourhood, and later steps convey fine tuning of the model with short radii which impact only one or zero surrounding nodes. Several choices for neighbourhood function shape can be used, the choice mostly affecting the earlier coarse organisation of the dataset. The size of $\sigma(t)$ should be large enough to organise across the whole of the map during early iterations, as if it is too small then data become clustered in local pockets across the map without global organisation \cite{Kohonen1990-sf}.

\subsubsection{Learning Rate.}
The learning rate, $\alpha(t)$, defines how the weight of each node within the neighbourhood is updated with each training iteration. It also decreases monotonically with each iteration, $t$, to allow early coarse organisation and later fine organisation of the data \cite{Kohonen2001-al}.

\subsubsection{Decay Function.}
The decay function defines how much the neighbourhood function radius and learning rate decrease with each training iteration. Decay functions may be any function of $t$ that allows $\sigma(t)$ and $\alpha(t)$ to decrease with increasing $t$ \cite{Kohonen2001-al}.

\subsubsection{Iteration Number.}
Due to the stochastic nature of SOM learning, many iteration steps are required to achieve convergence. There is no calculation available to determine ``enough'' iterations, but Kohonen suggests at least $500l$, where $l$ is the number of map nodes, to ensure good statistical accuracy \cite{Kohonen2001-al}.

\subsubsection{Computational Complexity.}
When dealing with larger maps and high dimensionality data, it may be important to optimise map parameters to reduce program runtime. The SOM algorithm demonstrates the following properties:
\begin{itemize}
    \item Linear with respect to the number of observations in the dataset, $n$
    \item Linear with respect to the number of dimensions of each observation, $k$
    \item Quadratic with respect to the number of nodes in the map lattice, $l$
\end{itemize}
Therefore, the overall complexity of the algorithm is $O(nkl^{2})$.

\subsubsection{SOM Error Metrics.}
No single metric can adequately describe the SOM method, and visual interpretation of the output is required to determine the use of the SOM for the question at hand \cite{Erwin1992-ot}. Two commonly used metrics to assess different aspects of the SOM are the quantisation and topographic errors.

The quantisation error is the average distance of each data point to its closest node within the lattice, and expresses how well the SOM represents the distribution of the input dataset. Increasing the number of lattice nodes decreases quantisation error, but risks overfitting the model, particularly when the ratio of data points to nodes becomes $\leq 1$ \cite{Ponmalai2019-vb}.

The topographic error denotes the proportion of input data vectors for which the best and second-best matching units are not adjacent in the map network, and therefore do not share a direct lateral connection \cite{Ponmalai2019-vb}. This metric reflects how well the SOM represents the topology of the input data.

\subsubsection{MiniSom and MySom.}
MiniSom \cite{MiniSom} is an open source Python package used for the computational SOM analysis. It is popular and versatile  \cite{Yuan2018-ym}.
We have developed the python module MySom, subclassing MiniSom, and adding methods for spectral normalisation and consistent SOM outputs \cite{West2021}. The source code for MySom is available at \url{github.com/thenakedcellist/prostate}.

\section{Our Data Sets}
The full dataset used consists of $n=284$ observations, 154 from a normal prostate cell line, PNT2-C2, and 130 from the malignant LNCaP prostate cell line. Each observation consists of a single Raman spectrum measurement comprising $k=1056$ wavenumber points. See also Section 2.2 for other data-specific details concerning the standard methods for preprocessing. The collected data comprise single-cell, point measurements within the cell nucleus randomly selected across the cell population. 

The sample data sets are tested for statistical robustness by converging the spectral mean, twice the standard error and the standard error statistical measures (see also methods in \cite{Rocha2021-za}). 

The Raman data with standard preprocessing (Section 2.2) were further intensity-normalised by dividing each column of the matrix by the Frobenius norm. No additional processing was performed; in particular, no binning was used to reduce dimensionality. All analyses in the main SOM experiment are first performed with blinded data (i.e., unlabelled observations) with the labelling not made available until after the initial analysis is performed. In this respect, the SOM method is fully unsupervised. 

\section{Preliminary Investigation}
\begin{figure}[tp]
\includegraphics[width=\textwidth]{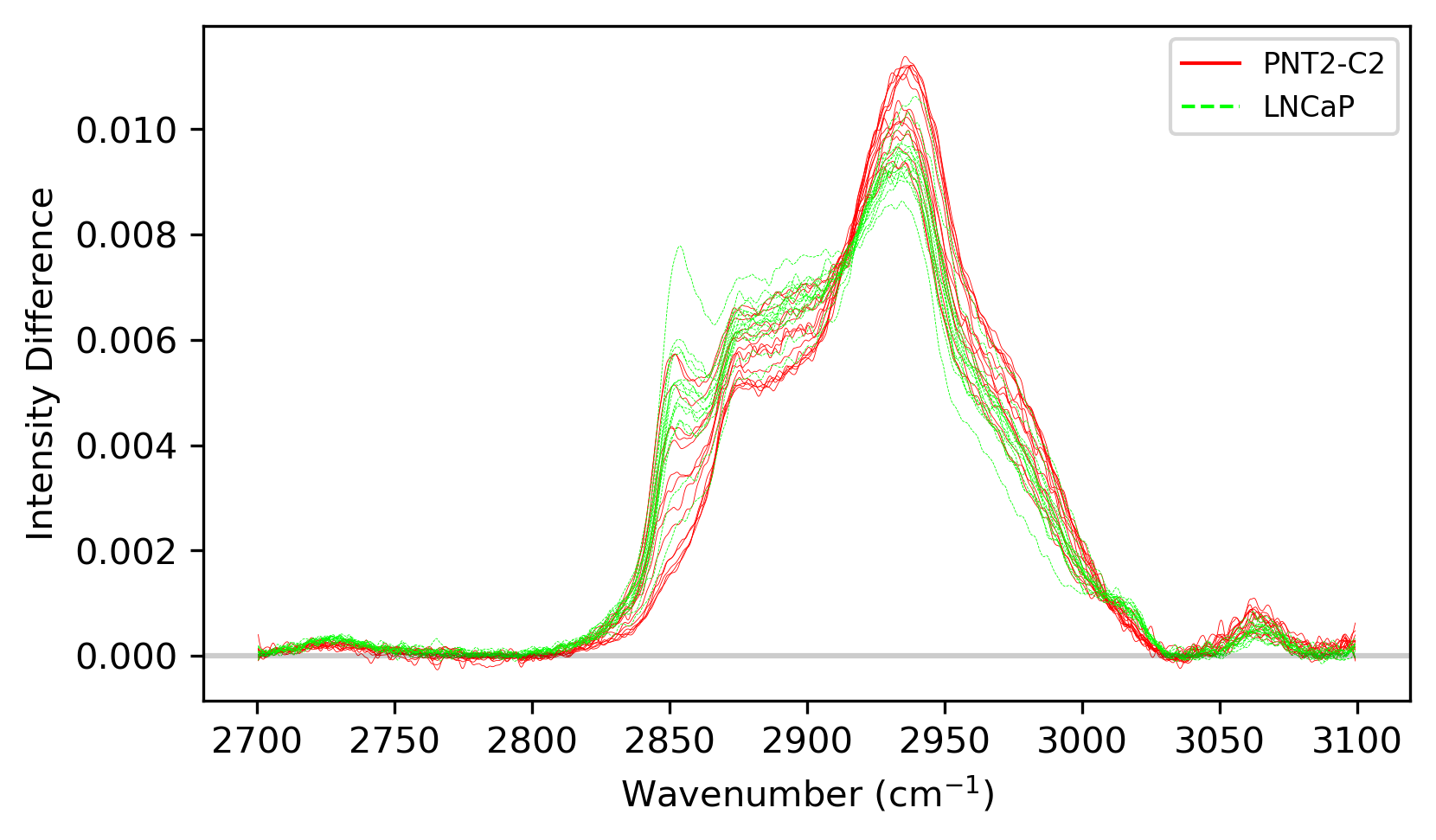}
\caption{Raman spectra generated from thirty cell nuclei: fifteen from the normal prostate cell-line (PNT2-C2), and fifteen from the cancerous prostate cell-line (LNCaP). The spectral intensity has been normalised and is displayed in arbitrary units.}
\label{small_spectra}
\end{figure}

The first experiment conducted was a feasibility study into the ability of a SOM to cluster Raman spectra. SOMs are not typically used with such high-dimensional, real-system data, hence, this experiment was performed to see if the method works without the need to further preprocess and dimension reduce the data, such as by data binning. The spectra for this initial study were obtained from 15 normal and 15 cancerous prostate cells selected randomly from the full, statistically-converged, population-level sampling. The resultant 30 spectra in Figure \ref{small_spectra} demonstrate some separability by eye between classes. However, at wavenumbers that correspond to known, putative biomarkers, such as at $\sim$2915~cm$^{-1}$ and $\sim$3015~cm$^{-1}$ \cite{Potcoava2014-tl}, the data from both clusters densely overlap and are indistinct. 

\begin{figure}[tp]
\centering
\includegraphics[width=0.9\textwidth]{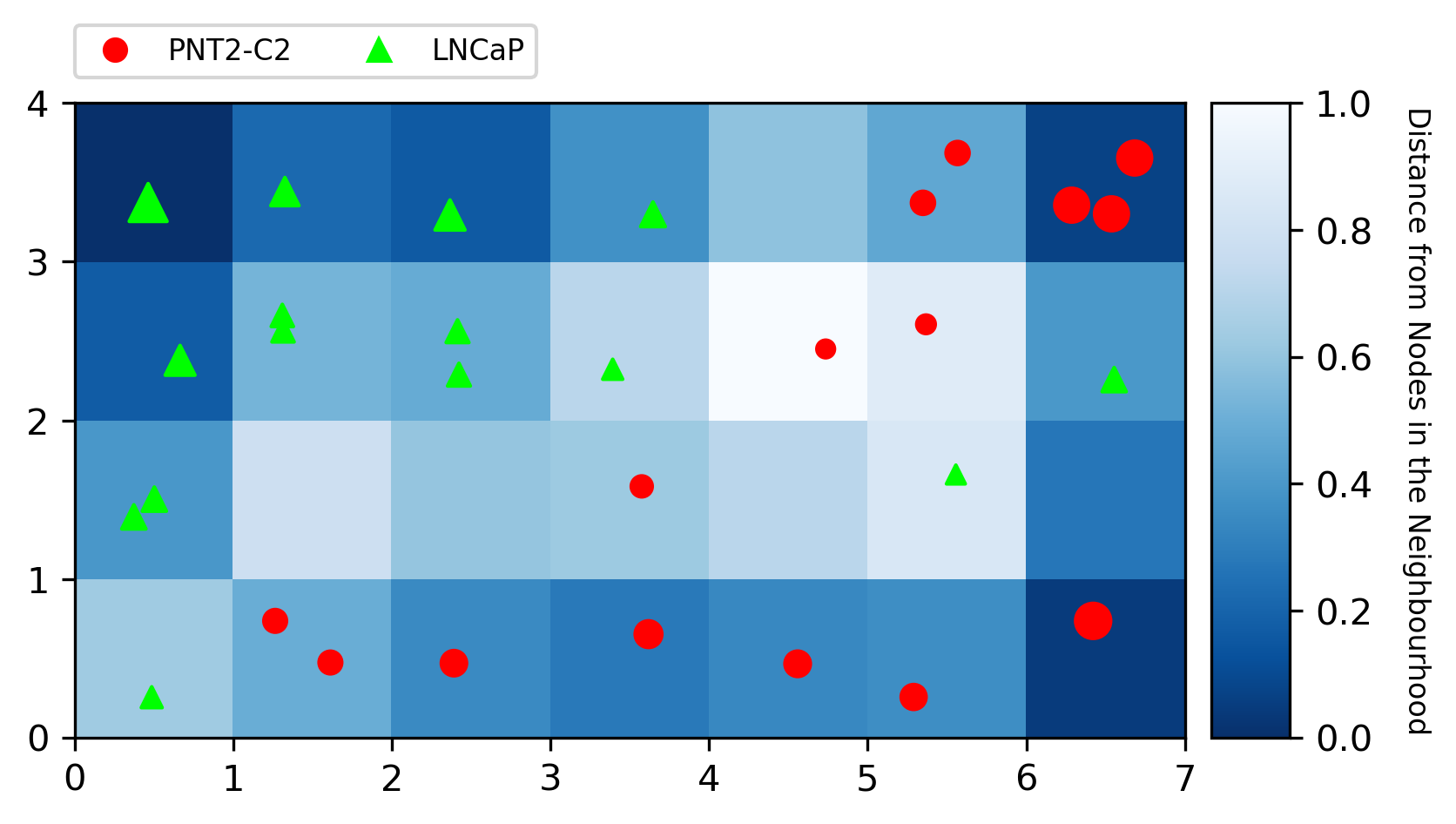}
\caption{SOM trained on the small Raman spectra dataset. A stripe of low density nodes across the central map separates the bulk of PNT2-C2 and LNCaP data.}
\label{small_som_scatter}
\end{figure}

The SOM in Figure \ref{small_som_scatter} shows separation of the two data clusters, clearer than the spectral representation in Figure \ref{small_spectra}. The blue colour of the SOM nodes represents the nodal density: darker blue nodes are more closely clustered together. Overlying each node are markers representing the Raman spectra mapping to that node. The size of a marker is proportional to the density of its node, such that the clustering of data within dense regions of the map is emphasised (large markers are closer together in the underlying space than are small markers).

The low-density (pale colour) stripe of nodes across the diagonal centre of the map from the lower left to upper right separates the PNT2-C2 cluster (right of map) from the LNCaP cluster (left of map). Two LNCaP observations are separated from the rest of this class at the extreme right of the map. This  result likely stems from the fine granular nature of this SOM as the number of SOM nodes ($n = 28$) is almost equal to the number of data points ($n = 30$). 

A full parameter optimisation was not employed with the initial dataset, as the number of observations is inherently related to SOM parameter choice, and the main experiment dataset consists of 284 observations.  However, the preliminary results are enough to give us confidence that the SOM method can be used to analyse Raman spectral data without the need for dimensionality reduction, only routine preprocessing as outlined in \cite{Rocha2021-za} plus intensity normalisation.

\section{Main Experiment}
The full dataset comprises 284 blinded samples from the normal PNT2-C2 prostate cell-line and from the cancerous LNCaP prostate cell-line. A parameter sweep is performed using the guidance outlined in Section \ref{params} to generate several iterations of SOM, and the resultant SOMs are then assessed based on error metrics and visual interpretation.

\subsubsection{SOM Analysis.}
\begin{figure}[tp]
\centering
\includegraphics[width=0.9\textwidth]{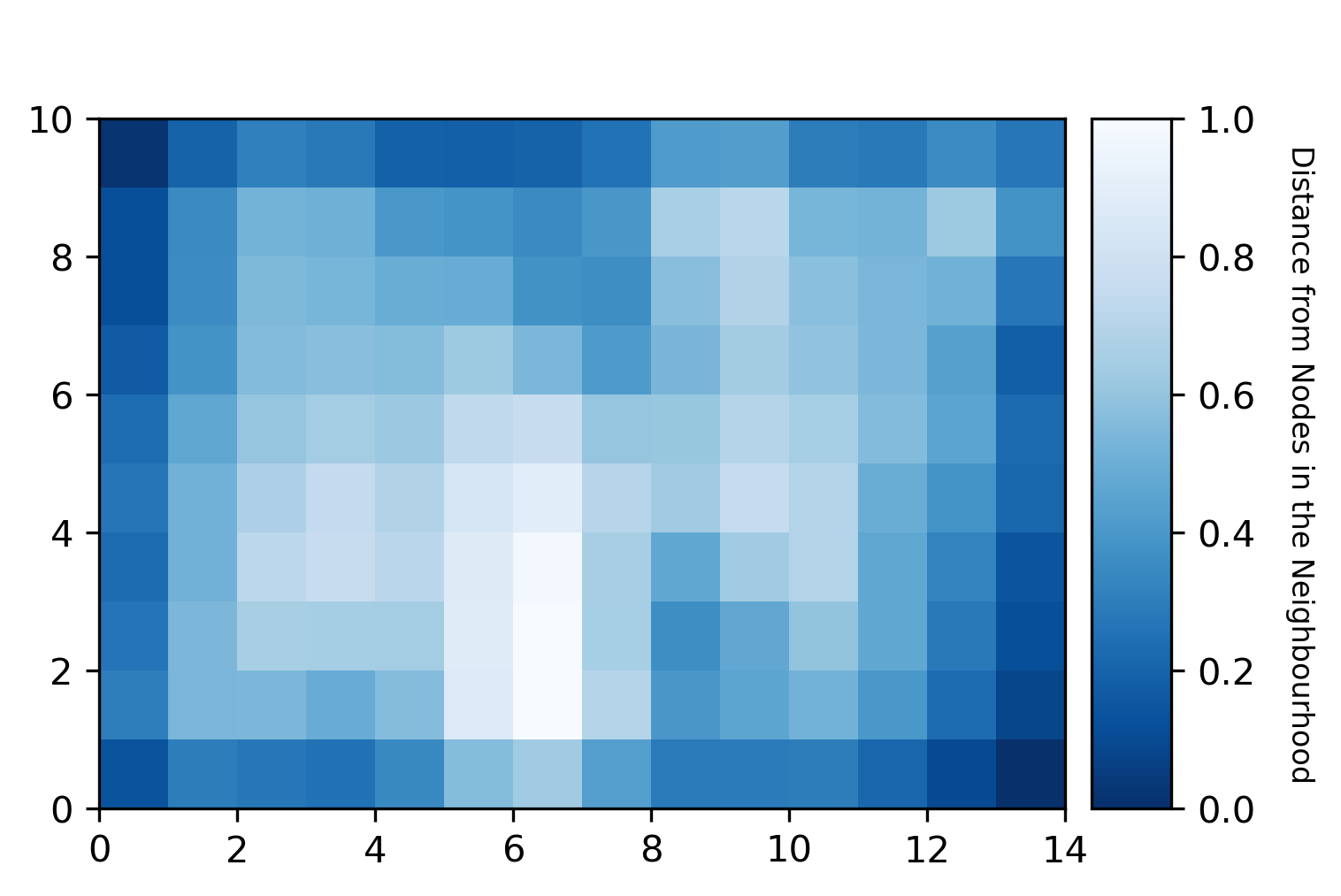}
\caption{SOM built with the full Raman spectra dataset. There is good spread of data across the entire map, with low density stripes from the bottom left to top right, and from the centre left to the centre bottom.}
\label{main_som}
\end{figure}

The SOM in Figure \ref{main_som} best displays clusters within the data, and supports Kohonen's suggestion of optimum training parameters, with a $\sigma(0)$ of at least half the map dimension and an $\alpha(0)$ of close to 1 \cite{Kohonen1990-sf,Kohonen2001-al}. However, the dimensions used for the map lattice are not in the ratio of the highest two eigenvalues in the dataset's autocorrelation matrix, as suggested by Kohonen to improve learning convergence during the early iterations. In our case, the ratio is $23:6$, which produces a long narrow output map that is difficult to interpret visually. Several pairs of map dimensions were tested to confirm stability of data clustering with different map configurations. Squarer maps are easier to interpret visually.  However, to ensure robust clustering with a rectangular lattice, the $x$ and $y$ dimensions should not be the same; breaking symmetry of the map in this manner ensures faster learning \cite{Kohonen2013-jv}. The number of iterations should be at least $500l$ where $l$ is the number of nodes, in this case $500 \times 14 \times 10 = 7 {\times} 10^4$. A value higher than this should be selected, but not too high as to render computational time excessive.
This testing leads to the following parameters used to produce the SOM in Figure \ref{main_som}:

\begin{itemize}
	\item Map dimensions: $14 \times 10$
	\item $\sigma(0)$: 3.0
	\item $\alpha(0)$: 0.75
    \item Maximum iteration number: $10 {\times} 10^5$
\end{itemize}

MiniSom uses a map lattice with edges rather than the more usual periodic boundary conditions. 
It calculates node distances as the sum of distances to all neighbouring nodes. Therefore, we would expect the calculated distances of edge and corner nodes to be smaller than those of the interior nodes, as they have fewer neighbours.
This effect is evidenced by the dark edge nodes and even darker corner nodes in figure~\ref{main_som}. Even taking this bias into account, however, there can be seen a medium- to low-density stripe across the map from the bottom left to the top right, and a very low-density stripe from the centre to the bottom of the map. These two stripes appear to separate the data into three clusters, rather than the expected two of normal (PNT2-C2) and cancer (LNCaP).

The uncovering of three clusters poses three distinct hypotheses:
\begin{itemize}
    \item Single subgroup hypothesis: one cluster contains normal observations, one contains cancerous observations, and the third contains observations from either the normal or disease state
    \item Mixed subgroup hypothesis: one cluster contains observations from normal or disease state, and the remaining two contain observations from both the disease and normal states
    \item External heterogeneity hypothesis: all three clusters contain data from normal and disease states, and clusters are based on other factors
\end{itemize}

\begin{figure}[tp]
\centering
\includegraphics[width=0.9\textwidth]{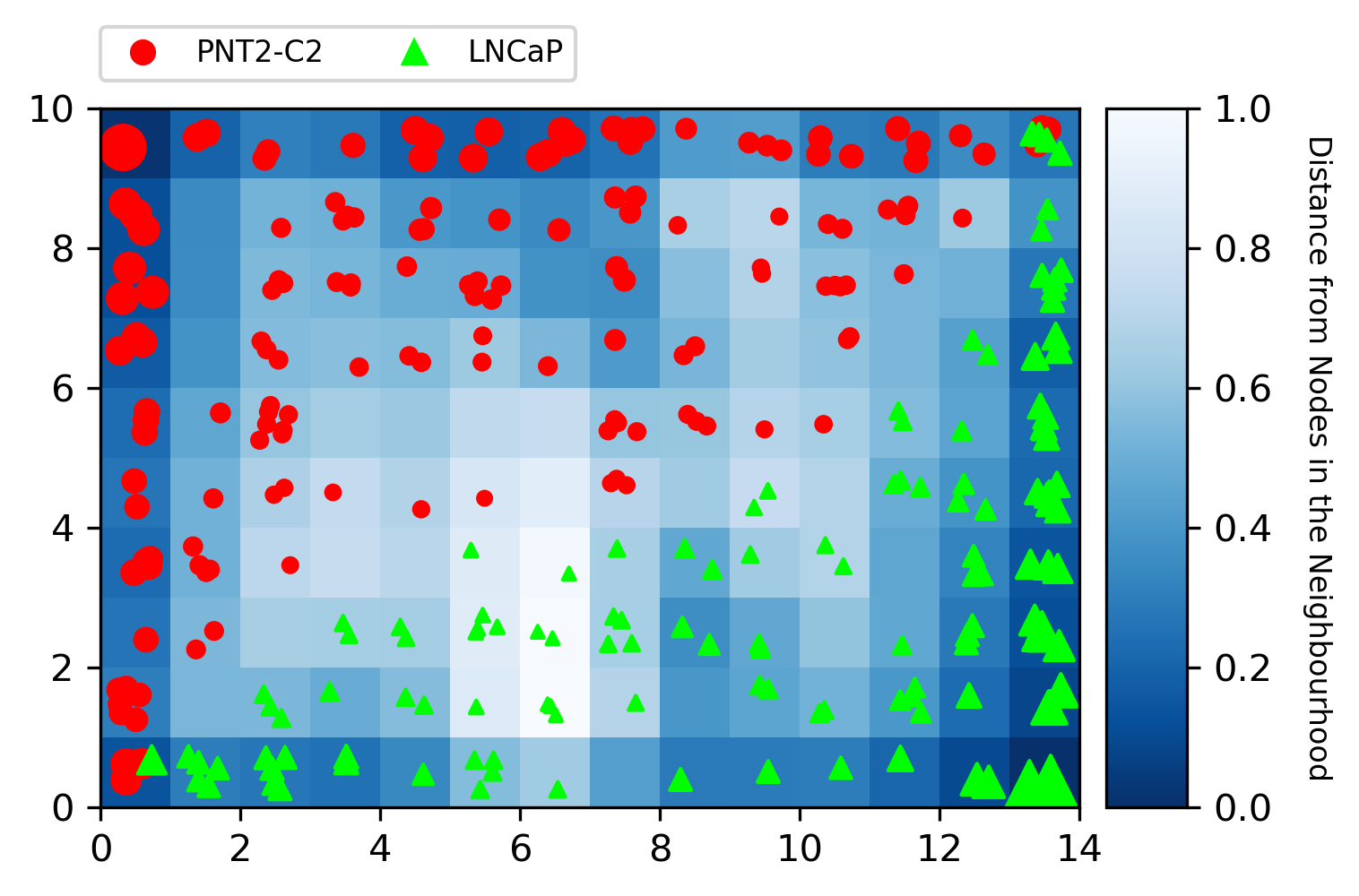}
\caption{SOM from Figure \ref{main_som} with observations mapping to each node overlaid. The PNT2-C2 observations are located to the top left, separated from the LNCaP observations by a diagonal, low-nodal density stripe. A vertical low-density stripe in the centre of the map divides the LNCaP group into two subclusters.}
\label{main_som_scatter}
\end{figure}

When the data are unblinded (Figure \ref{main_som_scatter}), they support the single subgroup hypothesis: one distinct cluster of observations from normal prostate (red), and two distinct clusters from the disease state (green). Hence, the SOM has successfully distinguished the normal from cancer cells.
Additionally it has revealed an unexpected, possible further subclustering of the cancer cells. To investigate the three clusters further, a threshold of nodal distance from neighbours was used to select the nodes which form clusters and the observations mapped to them. Table \ref{cluster_spectra_tab} outlines the clusters generated with a threshold distance of \textless 0.72.

\begin{table}[tp]
\centering
\begin{tabular}{ccccc}
\toprule
Cluster & Location & PNT2-C2 & LNCaP & Proportion\\
\midrule
A & Upper left & 121 & 0 & 0.426\\
B & Lower left & 0 & 18 & 0.063\\
C & Lower right & 0 & 73 & 0.257\\
\bottomrule
\end{tabular}
\caption{Table summarising the number of PNT2-C2 and LNCaP observations and the proportion of the total ($n = 284$) mapped to each cluster in Figure \ref{main_som_scatter}.}
\label{cluster_spectra_tab}
\end{table}

\subsubsection{Spectral Analysis.}
\begin{figure}[tp]
\includegraphics[width=\textwidth]{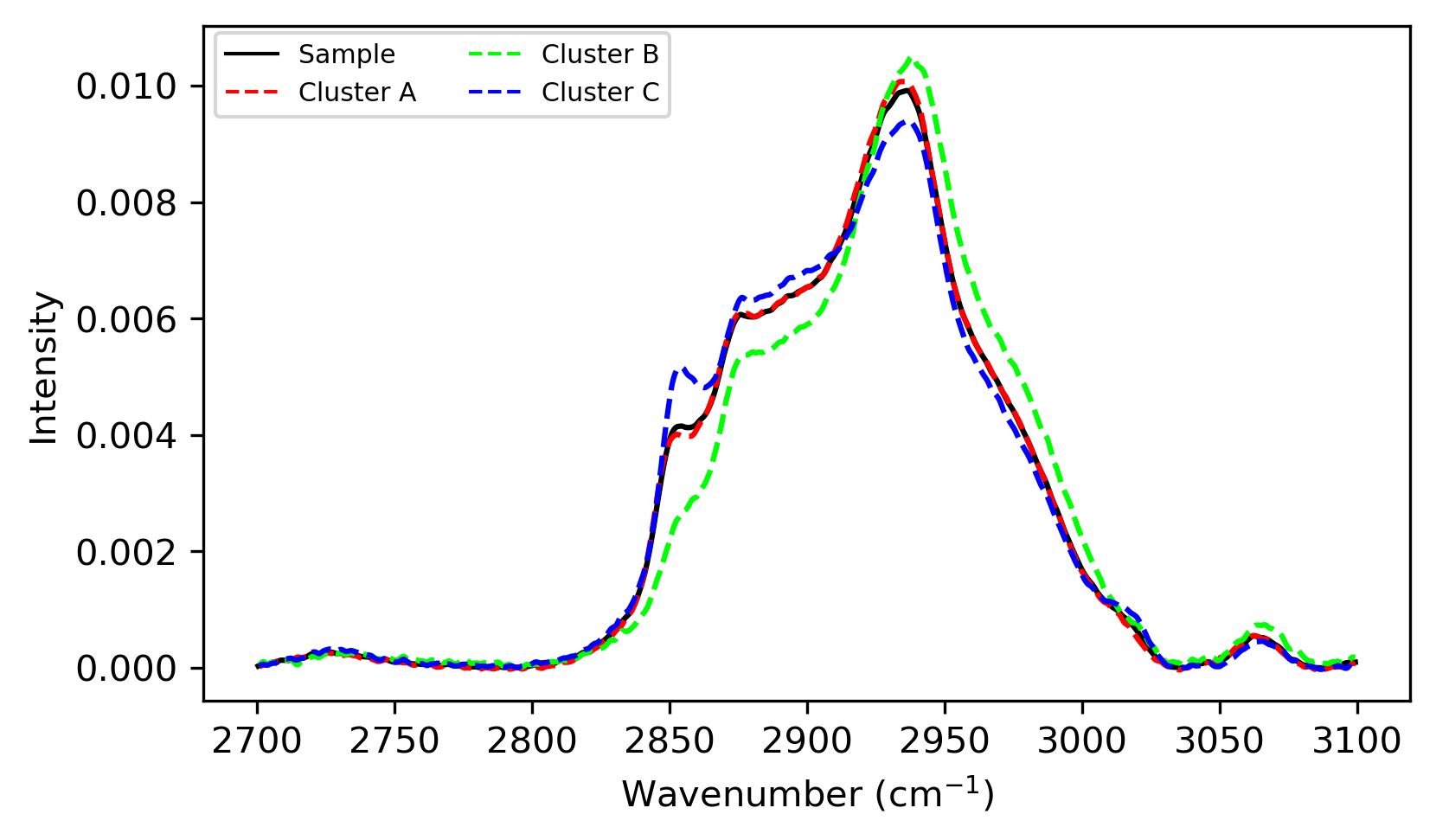}
\caption{Mean Raman spectra of the full dataset and each cluster.}
\label{cluster_mean_spectra}	
\end{figure}

The full dataset ($n = 284$) is used to build the SOM in Figure \ref{main_som}. Only the Raman spectra observations mapping to the cluster centres as outlined in Table \ref{cluster_spectra_tab} are used for spectral analysis. Due to the edge and corner effects of MiniSom, the weights of these boundary nodes are not used to determine the clusters, but observations mapping to boundary nodes are included where these nodes are in direct connection with the nodes used to determine the clusters. This corresponds to three quarters of the observations in the dataset, and the mean spectra are presented in Figure \ref{cluster_mean_spectra}.

Cluster A (normal) follows a similar spectral pattern to the whole dataset. Cluster B (cancer) has lower intensity at the 2850~cm$^{-1}$ to 2930~cm$^{-1}$ segment (consistent with lower saturated fatty acids), and slightly higher intensity from the 2940~cm$^{-1}$ peak to 3010~cm$^{-1}$ (consistent with higher unsaturated fatty acids). Cluster C (cancer) follows a similar pattern to the whole dataset, with a noticeably higher intensity shoulder at 2850~cm$^{-1}$ (higher saturated fatty acids). The general trend shows slightly higher intensity from 2870~cm$^{-1}$ to 2920~cm$^{-1}$, and then slightly less intensity from 2930~cm$^{-1}$ to 3010~cm$^{-1}$. 

\begin{figure}[tp]
\includegraphics[width=\textwidth]{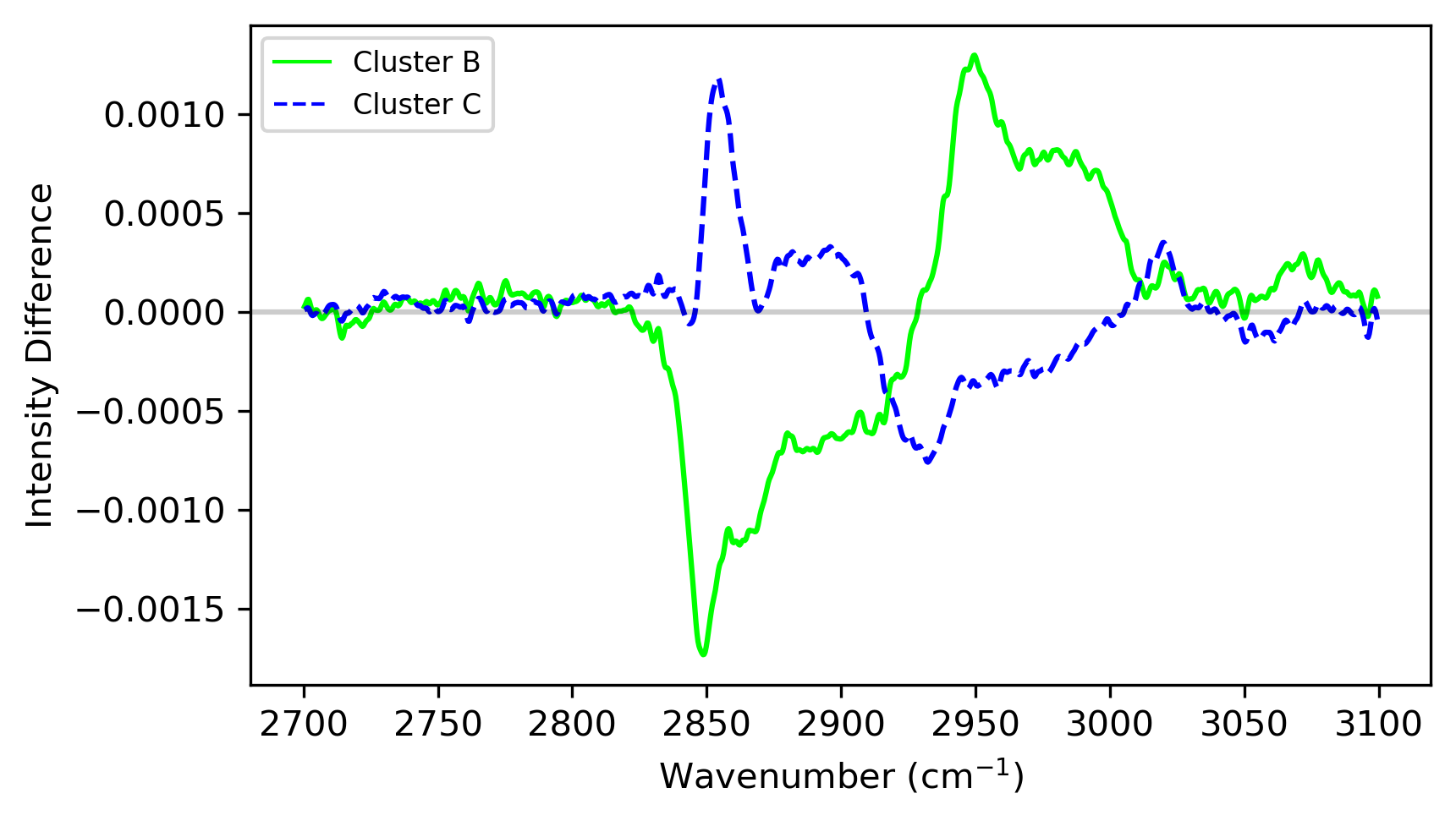}
\caption{Spectral subtractions of the two cancer clusters (B and C) from the normal cluster (A).
}
\label{spectral_subtraction}
\end{figure}

Mean spectral subtractions of the two cancer subclusters relative to the PNT2-C2 normal baseline quantitatively show the spectral difference results (Figure \ref{spectral_subtraction}). The transition from saturated fatty acid dominance to unsaturated fatty acid dominance as a function of increasing wavenumber in Raman spectra within the 2700--3100~cm$^{-1}$ spectral region has been reported in Refs.~\cite{Movasaghi2007-lz,Talari2015-va}. Further analysis of the SOM outputs, including statistical assessment of these data subsets, could enable biologically plausible reasons for the different lipid signals within the cancerous cells to be conjectured. Lipids can be used as intracellular signalling molecules \cite{Wymann2008-kz} and it is known that obesity is associated with worse outcomes for patients with prostate cancer \cite{Saha2023-ps}. Together, these findings support further work into the investigation of Raman spectral characteristics of different cell types, and the cluster analysis of these data using SOMs.

\section{Summary and Conclusions}
This work has demonstrated, as proof-of-feasibility, the ability of Kohonen self-organising maps (SOMs) to cluster complex biological data. The data used were of high dimensionality with no dimension reduction being performed. To our knowledge, an unsupervised SOM method has not been used before on such high-dimensional, real system data. The limit of the experimental resolution regarding SOM application lies with the Raman spectroscopy output data together with its standard pre-processing (baseline subtraction, total area normalisation and interpolation). Spectral data are then intensity normalised for SOM input, but do not require dimensionality reduction to yield meaningful results. 

The application of the unsupervised SOM in this work has shown its ability to substratify prostate cancer, live-cell Raman data, with the results indicating relative differences in saturated and unsaturated fatty acid content between the two cancer subclusters. Future work will involve further analysis of the SOM outputs and deeper focus on the biological relevance of these results. The Mini\-Som package provides a flexible tool for easily building of SOMs. However, its limitation concerning edge effects should be noted, particularly when using small SOM maps. 

\subsection*{Acknowledgements}
The Raman data were generated under Dr Y.\ Hancock's supervision in the School of Physics, Engineering and Technology at the University of York by Mr Charles Kershaw (PNT2-C2 cell line), and Dr Marcus Cameron (LNCaP cell line). The PNT2-C2 and LNCaP cell lines were cultured in Professor Norman Maitland’s laboratory in the Cancer Research Unit, Department of Biology at the University of York, with direct assistance from Dr Fiona Frame. The methods associated with the Raman data generation, including the biological sample preparation, can be found in 
\cite{Kershaw2017} for the PNT2-C2 cell line, and 
\cite{Cameron2021} for the LNCaP cell line.

\paragraph{Data availability:}
Datasets analysed for this study are available by request to Y. Hancock (y.hancock@york.ac.uk) from the University of York repository (website to be provided).

\bibliography{bibliography}

\end{document}